\documentclass[10pt,hyper]{JHEP}

\usepackage{amsmath}
\usepackage{amsfonts}

\newcommand {\tr}   {\mathop {\rm Tr}}
\newcommand {\pexp} {\mathop {\rm Pexp}}
\renewcommand {\exp} {\mathop {\rm exp}}
\newcommand {\ctg} {\mathop {\rm ctg}\nolimits}

\renewcommand {\d}    {\partial}
\newcommand {\dt}   {\partial_{\tau}}
\newcommand {\dn}   {\partial_{n}}
\newcommand {\Fi}   {\Phi}
\renewcommand {\phi} {\varphi}
\newcommand {\D}    {\mathcal D}
\newcommand {\G} {\mathop {\Gamma}}
\newcommand {\Op} {\mathcal O}
\newcommand {\s}    {\sigma}
\newcommand {\f}  {\frac}
\renewcommand {\ap} {{\alpha'}}
\newcommand {\e}  {\varepsilon}
\newcommand {\de} {\delta}
\newcommand {\wt} {\widetilde}

\newcommand {\la} {\left \langle}
\newcommand {\ra} {\right \rangle}

\newcommand {\lb} {\left (}
\newcommand {\rb} {\right )}

\newcommand {\pdt}[1] {{\rm P}[d^{#1}t]}
\renewcommand {\t}{\dot}
\newcommand {\st}{\acute}
\newcommand {\PP}{{\rm p} + 1}
\newcommand {\zz}{\mathcal{Z}}

\newcommand {\bft} {\begin{footnote}}
\newcommand {\eft} {\end{footnote}}

\newcommand {\bfoot} {\begin{footnote}}
\newcommand {\efoot} {\end{footnote}}

\newcommand {\be} {\begin{equation}}
\newcommand {\ee} {\end{equation}}

\newcommand {\req} {\stackrel{\rm reg}{=}}
\newcommand {\FiPic} {\begin{picture}(23,10)
\put(0,3){\line(1,0){21}} \put(7,-2){\line(0,1){10}}
\put(14,-2){\line(0,1){10}}
\end{picture}}

\title{On non-abelian low energy effective action for D-branes}

\author{Vasily Pestun\\
ITEP, Moscow, Russia.\\
E-mail: \email{pestun@gate.itep.ru}}

\preprint{ITEP-TH-52/01\\ \hepth{0110092}}

\abstract{
Connection between the partition function for the $2D$
sigma model with boundary pertubations and the low energy
effective action for massless fields in the open string theory
is discussed. In the non-abelian case with a stack of $N$ D-branes,
the terms up to the order of $\ap^3$ are found.
}

\keywords{effective action, D-brane}

\begin{document}

\section {Introduction}
Open string theory can be considered as a field theory with
infinite number of fields, corresponding to different string
excitations. Although string theory tells us how to calculate
scattering amplitudes~\cite{StrTh} for these fields, we still do not know the
corresponding field theory action.
In the low energy limit one may be interested in the effective action for massless fields (with
massive ones integrated out). All amplitudes calculated using this
effective action should coincide with those obtained in string theory. As
a first approximation, one may try to find some "classical"
effective action, which is only demanded to reproduce the tree
diagrams. One way to construct it is to consider the generating
functional for the connected tree diagrams, which can be written
as the partition function for an appropriate $2$-$D$ sigma model on
the string world-sheet \cite{Tseytlin_effective,OSFT}
\begin{equation}
Z[J]= \int \D X[\sigma] e^{-S_{\rm free}[X]}e^{-S_{\rm
int}[X,J]}
\end{equation}
Here $\sigma^a$ are coordinates on the flat unit
disk $D$, $S_{\rm free}[X]$ is the free action
\begin{equation}
S_{\rm free}[X]=\f 1 {4\pi\ap}\int_{D} {\d_a X_{\mu} \d^a X^{\mu} } \,
d^2\sigma \label{sfree} \end{equation}
and $e^{-S_{\rm int}[X,J]}$ is some
boundary perturbation
\begin{equation}
e^{-S_{\rm int}[X,J]}= \tr \pexp \lb
\oint_{\d D} \Op_i[X] J_i d\tau \rb
\end{equation}
with appropriate boundary vertex operators  $\Op_i$. It is easy to see that in the critical
dimension and for the conformal boundary perturbation $Z[J]$ is,
indeed, a generating functional for connected tree diagrams (without
external legs).

Now suppose that we have a second quantized
field theory with some action $S[\phi]$, then the generating functional of all
connected diagrams is
\begin{equation}
e^{-W[J]} = \int \D \phi e^{-S[\phi]-J\phi}
\end{equation}
If interested in the connected tree
diagrams (i.e. the semi-classical approximation), one should evaluate
integrand in the saddle point
\begin{equation}
W_{\rm tree}[J] = S[\phi_c]+J\phi_c, \qquad \f {\de S} {\de \phi_c} = - J \label {wct}
\end{equation}
Thus, the generating functional for the connected tree
diagrams and the action are related by the Legendre transformation.

The generating functional for the connected tree diagrams without external legs
is  $W_{\rm tree}[KJ]$, where $K$ is the kinetic energy operator in the action
\begin{equation}
S[\phi] = \f 1 2 \phi K \phi + V[\phi]. \label{action}
\end{equation}

The $2$-$D$ sigma model partition function $Z[J]$ is also the generating functional
for the connected tree diagrams without legs, thus we require that $Z[J]=W_{\rm tree}[KJ]$.
Let us see, how $Z[J]$ and the action $S[\phi]$ are
related pertubatively in the leading order.
From (\ref{wct}) and (\ref{action}) we have
\begin{equation}
K\phi_c + \f {\delta V}
{\delta \phi_c} = -J
\end{equation}
Then, pertubatively,
\begin{align}
\phi_c[J] = -K^{-1}J - K^{-1} \f {\de V} {\de \phi} [K^{-1}J] + \dots\\
\end{align}
Substituting $\phi_c[KJ]$ into (\ref{wct}), one gets in the leading
order
\begin{equation}
Z[J] = W[KJ] = -\f 1 2 J K J + V[J] + \f 1 2 \f {\de V}
{\de \phi} K^{-1} \f {\de V} {\de \phi} + \dots \label{zrel}
\end{equation}
$Z[J]$ is the generating functional for all the connected tree diagrams and
is contributed by the following term $V[J]$ describing 1-vertex diagrams, by the term
$\f 1 2 \f {\de V} {\de \phi} K^{-1} \f {\de V} {\de \phi}$
corresponding to the diagrams with one internal line, and so on.
Note that $Z[J]$ has poles ($K^{-1}$) when any propagating particle inside some
diagrams lies on-shell (i.e. has $p^2=0$ for massless fields).

To find $S[\phi]$ from $W[J]=Z[\wt{J}],\quad \wt{J} \equiv K^{-1} J$,
one needs to make the inverse Legendre transformation. When
\begin{equation}
Z[\wt{J}] = -\f 1 2 \wt{J} K \wt{J} + \wt{Z}[\wt{J}]
\end{equation}
one has
\begin{equation}
S[\phi] = \left . Z[\wt{J}] - Z[\wt {J}]\f {\delta Z} {\delta \wt{J}} \right
|_{ \f {\delta Z} {\delta \wt{J}} = K\phi} \label{letrans}
\end{equation}
$\wt{J}$ should be expressed in terms of $\phi$. One can do it pertubatively
\begin{equation}
\wt{J}=-\phi +  K^{-1} \f {\de \wt{Z}} {\de \wt{J}}[\wt{J}(\phi)] =
-\phi +  K^{-1} \f {\de \wt{Z}} {\de \wt{J}}[-\phi + \dots] \label{phirenorm}
\end{equation}

We are going to evaluate $2D$-sigma model partition function
pertubatively in momenta
\bfoot
{i.e. in derivatives in  the coordinate representation.}
\efoot
(but it will allow us to get in some
cases a nonpertubative in the field strength expression).

For instance, let us consider, the simplest case:
potential for the constant tachyon field with the boundary coupling $\oint_{\d D} \wt{T} \d \tau $:
\begin{equation}
Z[\wt{T}] = \la e^{-\f 1 {2\pi }\oint_{\d D} \wt{T} \d \tau } \ra = e^{-\wt{T}}
\end{equation}
Making the Legendre transformation (\ref{letrans}) one obtains the exact tachyon potential \cite{Gerasimov_tachyon}
\begin{equation}
S[\wt{T}] = e^{-\wt{T}}(1+\wt{T})
\end{equation}
and, after the field redefinition $e^{-\wt{T}}=T$, $\wt{T}=\log (-T)$, one obtains
\begin{equation}
S[T] = T \log \lb \f {-T} e \rb
\end{equation}
We see that in this particular case (\ref{letrans}) works well.

Our next goal is to obtain the effective action for the massless fields living on a stack
of $N$ parallel  D-branes. We expect  the partition function suffers from divergences
in this case, since $K^{-1} \sim \f 1 {p^2}$ has pole at $p^2=0$. In terms
of the $2D$ sigma model, as we will see, it will be a logarithmic divergence
at small distances due to massless exchanges and a linear divergence due to the tachyon exchange.
Thus, the partition function needs some regularization and renormalization. The regularization
can be done by modifying $2D$-propagators, which is equivalent to the $2D$ world-sheet cut-off
at small distances and at the same time it implies an IR-cutoff in the target space-time.
This IR space-time divergence arises when the string world-sheet can be represented as (i.e.
the unit disk conformally transformed to) a long strip which connects the initial and the final states.
The length $L$ of this strip is proportional to $\ln \f 1 \e$, where $\e$ characterize how
close may be the points on the unit disk.
The expression for the propagator from the first quantized field
theory is
\begin{equation} \int_{0}^{\infty} e^{- \ap (p^2 + m^2)L} \, dL = \f
1 {\ap(p^2 + m^2)} \label{prop}
\end{equation}
In terms of  string theory, $L$ is connected with parameterization
of the moduli space of string surface in the
neighborhood of its degeneracy.  As was stated before, the string surface
becomes degenerate when the points $t_i$, in which the vertex operators have been
inserted, approach each other: $t_i \to t_j$.
In these terms,
\begin{equation}
L \sim \ln \lb \f 1 {|t_i-t_j|}\rb
\end{equation}
We work with the perturbative expansion over momentum at the point $p=0$,
therefore, (\ref{prop}) diverges as~$L$, when $m^2=0$ and as $e^{-\ap
m^2 L}$, when $\ap m^2 < 0$. The $\e$-regularization, which we are going to use,
effectively bounds $\min|t_i-t_j|\simeq \e$, i.e. $\max
L \simeq \ln (\f 1 \e)$. Hence, we shall get the linear divergence $\f 1
\e$, when $\ap m^2 = -1$ (that is the tachyon exchange),
and the logarithmic one $\ln \f 1 \e$, when $m^2 = 0$ (for the massless particle exchange).
In the string amplitude non-perturbative over momentum the first divergence is hidden,
since an analytical continuation in momentum is used~\cite{StrTh},
(at $p=0$ after the analytical continuation the propagator for the tachyon is $-\f 1 \ap$, see below),
while the second, logarithmic divergence describes the pole $\f 1 {p^2}$ at $p=0$.

In the superstring calculation, when the tachyon is decoupled, only logarithmic
divergences (i.e. massless particle exchanges) survive. The way to get the effective
action from the partition function is, indeed, the renormalization procedure
\cite{tseytlin,ren} for the $2D$-sigma model with boundary couplings (which are fields in the target space-time), since one should
substitute
\begin{equation}
\left . K^{-1} \right | _{p^2 = 0}  = \left . \f 1
{(2\ap \pi)^2}  \f 1 {p^2} \right | _{p^2 = 0 }  \leftrightarrow
\f 1 {2\ap \pi^2} \ln \f 1 \e
\end{equation}
in (\ref{phirenorm},\ref{letrans})
\begin{equation}
S[\phi] = \left . Z[\wt{\phi}] - Z[\wt {\phi}]\f {\delta Z} {\delta \wt{\phi}} \right
|_{ \f {\delta Z} {\delta \wt{\phi}} = K\phi}
\end{equation}
\begin{equation}
\wt{\phi}=-\phi +  \f 1 {2\ap \pi^2} \ln \f 1 \e \f {\de \wt{Z}} {\de \wt{\phi}}[\wt{\phi}(\phi)] =
-\phi +  \f 1 {2\ap \pi^2} \ln \f 1 \e \f {\de \wt{Z}} {\de \wt{\phi}}[-\phi + \dots]
\end{equation}
Note that the renormalization procedure is not unique, since one can redefine $K^{-1}$
\begin{equation}
\left . K^{-1} \right | _{p^2 = 0}  = \left . \f 1
{(2\ap \pi)^2}  \f 1 {p^2} \right | _{p^2 = 0 }  \leftrightarrow
\f 1 {2\ap \pi^2} (\ln \f 1 \e + {\rm const})
\end{equation}
The resulting ambiguity in the effective action (the renormalized partition function)
is indeed due to the field redefinition ambiguity.

In the bosonic case, the situation is more difficult due to
presence of the tachyon coupling to massless fields. We have two ways
to find the action for the massless fields. First, one can include the tachyon
boundary coupling into the partition function, then to
construct the action including both the tachyon field and the massless fields,
and then to exclude the tachyon by the equations of motion. The second way is
to calculate the partition function with only the massless fields
turned on. In this case the tachyon field is already
integrated out, but before constructing the action
from the partition function (i.e. Legendre transformation or
renormalization of logarithmic divergences), one should analytically
continue amplitudes in momentum in order to remove the linear divergence emerged
due to the tachyon exchange.

\section{Stack of $N$ D-branes}

Now we are going to find the low energy effective action
for the massless fields living on a stack of $N$ D$p$-branes \cite{Polchinsky}.
In this case, the partition function is
\begin{equation} Z[A,\Fi] = \la \tr \pexp \left\{\oint \left(A_k
\dt X^k + i \Fi_i \dn X^i \, \right)d\tau \right\} \ra\label {pf}
\end{equation}
Here the averaging is done with $e^{-S_{free}[X]}$ (\ref{sfree}) and the functional integration
is performed over the world-sheet embedding into the space-time with the following boundary
conditions
\begin{align}
 &\st{X}^k  \equiv \dn X^k = 0 & & k=0  ,\cdots,p  \label {boundt}\\
 &\t{X}^i \equiv \dt X^i =  0 & &i=p+1,\cdots,D-1   \label {boundn}
\end{align}

The coordinates $X^{k}$ run along D-brane, $X^i$ run in perpendicular directions,
$\dt$ is a tangent derivative along the disk boundary, and $\dn$ is a normal derivative.
These conditions mean that open strings end on the flat $p$-dimensional hyperplane,
that is D$p$-brane. The vector field $A_k$ and $D-p-1$ scalars $\Phi^i$ in (\ref{pf})
describe the massless sector of open strings with the boundary conditions (\ref{boundt},\ref{boundn}).
In the case of a single D$p$-brane, the gauge group is $U(1)$, the fields are abelian,
and the resulting effective action can be found in all orders in the constant field
strength $F_{kl}$ (it is DBI-action \cite{Tseytlin_review}).
However, in the case of a stack of $N$ D$p$-brane,  the fields $A_k$ and $\Phi_i$ become
non-abelian $U(N)$ fields: they carry additional $U(N)$ Chan-Paton indices \cite{Gerasimov_nonab}.
\bfoot
{$N\times N$ indices describe pairs of N branes that the ends of the
open string are attached to.}
\efoot
In the non-abelian case there is no unambiguous way to extract from the effective action
the part that does not depend on the derivatives of the field strength \cite{Tseytlin_review_nonabelian}.
\bfoot{Commutators of the field strength can be rewritten through covariant derivatives.}
\efoot
Therefore, we are going to get the first terms of the non-abelian action in the expansion in $\ap$.

Let us illustrate that the $2D$ sigma model partition function is, indeed, the generating
functional for the connected tree diagrams.
For example, for the field $A$ we have
\bfoot{$\oint \pdt{n} \equiv
\int \limits_{0}^{2\pi} dt_n  \int \limits_{0}^{t_n}
dt_{n-1}\cdots \int \limits_{0}^{t_2} dt_{1}$}\efoot
\begin{multline}
Z[A] = \int \D X \exp{\left\{ -\f 1 {4\pi\ap}\int {\d_a X_{\mu}
\d^a X^{\mu} } \, d^2\sigma \right\}} \tr \pexp \left\{\oint
\left(A_k \dt X^k \, \right)d\tau \right\} = \\
 =\sum_{n=0}^{\infty} \int \D X \exp{\left\{ -\f 1 {4\pi\ap}\int {\d_a X_{\mu}
\d^a X^{\mu} } \, d^2\sigma \right\}} \oint \pdt{n}
\prod_{\alpha=1}^{n} A_{k_{\alpha}} (X(t_{\alpha}))
{\t{X}^{k_{\alpha}}(t_{\alpha})}
\end{multline}
Then,  let us use the Fourier expansion for $A_k(X)$
$$A_k(X) =\sum_{q} A_k(q) e^{iqX}$$
We shall decompose functions $X(\s)$ to zero modes and
deviations "orthogonal" to them
\begin{align}
 X^{\mu}(\s) = X^{\mu} + x^{\mu}(\s)\\
 \int d^2 \sigma \, x^{\mu}(\sigma) = 0
\end{align}
After that, we can explicitly extract the integration over
$X^{\mu}$ from the functional integral
\begin{multline}
Z[A] =  \sum_{n=0}^{\infty}  \sum_{q_{\alpha}} \int d^D
Xe^{\sum_{\alpha=1}^{n} iq_{\alpha}X_0} \int \D x e^{-S_{\rm free}[x]}
\oint \pdt{n}
\prod_{\alpha=1}^{n} A_{k_{\alpha}}(q_{\alpha}) e^{iq_{\alpha} x(t_{\alpha} ) } {\t{x}^{k_{\alpha}}(t_{\alpha})} =\\
=\sum_{n=0}^{\infty}  \sum_{q_{\alpha}} (2 \pi)^D \delta
\left(\sum_{\alpha=1}^{n} q_{\alpha} \right) \oint \pdt{n}
\prod_{\alpha=1}^{n} A_{k_{\alpha}}(q_{\alpha})  \la
\prod_{\alpha=1}^{n} e^{iq_{\alpha} x(t_{\alpha} ) }
{\t{x}^{k_{\alpha}}(t_{\alpha})} \ra
\end{multline}
The angle brackets denote the averaging with the weight of
$\exp(-S_{\rm free }[x])$ (\ref{sfree})
\begin{equation}
\langle \mathcal{F}[x] \rangle = \int \D x
\exp{\left\{ -\f 1 2 \int {x^{\mu} G^{-1}_{\mu\nu} x^{\nu} }  \,
d^2\sigma \right\} \mathcal{F}[x]}
\end{equation}
Functional measure for the Gaussian integral is defined so that
\begin{equation}
\langle 1 \rangle = 1
\end{equation}
Thus, $Z[A]$ is equal to 1 plus the sum of string amplitudes with the
number of external legs varying from $1$ to~$\infty$.
\begin{footnote}
{The expression in the angle brackets is the product of vertex
operators inserted to the string boundary. $\int \limits_{0}^{2\pi} dt_n  \int \limits_{0}^{t_n}
dt_{n-1}\cdots \int \limits_{0}^{t_2} dt_{1}$ is the path ordered
integral over the points where these vertex operators are inserted.
($t_{\alpha}$ parameterize the disk boundary.) $\delta
\left(\sum_{\alpha=1}^{n} q_{\alpha} \right)$ is due to the conservation of
momentum. Thus, the $n$-th term is exactly the expression for the
string scattering amplitude.}
\end{footnote}

\section{The partition function}
Let us start  calculating the partition function (\ref{pf}) $Z[\Fi,A]$, with the case of $A$ equal to zero.
In this case, the partition function~(\ref{pf}) reduces to
\begin{equation}
Z[\Fi] = \la \tr \pexp \left\{\oint \left(i \Fi_i \dn X^i \,
\right)d\tau    \right\} \ra \label{Def_Z_FI}
\end{equation}
Indeed, it is enough to calculate only this part of the partition function, since the remaining part,
the dependence on $A(X)$, can be restored by T-duality: one has to substitute $\nabla_i=\d_i+A_i$ instead of $\Fi_i$.
Unfortunately, there is no simple way to take the
integral, since in the non-abelian case, due to the $P$-ordering, the functional integral is not Gaussian.
We shall calculate it by expansion in series of $\pexp$
\begin{equation}
\begin{aligned}
\la \tr \pexp \left\{\oint \left(i \Fi_i \dn X^i \, \right)d\tau    \right\} \ra = 1 + Z_2[\Fi] + o[\Fi^2] \\
Z_2[\Fi]=- \tr \int d^{p+1} X\la \oint \pdt{2} \Fi_i(X+ x(t_1)) \Fi_j(X+ x(t_2)) \st{x}^i(t_1) \st{x}^j(t_2)  \ra = \\
= - \sum_{q_1,q_2} (2\pi)^{p} \delta (q_1 + q_2) \tr \Fi_i(q_1)
\Fi_j(q_2) \la \oint \pdt{2} e^{iq_1 x (t_1)} e^{iq_2 x(t_2)}
\st{x}^i(t_1) \st{x}^j(t_2) \ra
\end{aligned}
\end{equation}
After performing the Gaussian integration (see the Appendix for more
details) we get (using the regularization by analytical continuation
in momentum):
\begin{equation}
Z_2[\Fi]= -\sum_{q} (2\pi)^{p+1} \, \tr \Fi_i(q)
\Fi_i(-q) \ap \pi^{3/2}  4^{-\ap q^2 } \f {\G(-\ap q^2 - 1/2)}
{\G(-\ap q^2)}
\end{equation}
This expression is not perturbative in $q$, however, when $q$ is nearby zero it reduces to
\begin{equation}
Z_2[\Fi]=  -2 \ap^2 \pi^2
\sum_{q} (2\pi)^{p+1} q^2 \tr \Fi_i(q) \Fi_i(-q) =-2\ap^2 \pi^2
\int d^{p+1} X \, \tr \d_{k} \Fi_i \d_{k} \Fi_i \, \label{fip}
\end{equation}

Let us compare it with the initially pertubative in momentum
\bfoot{or in derivatives in the coordinate representation}\efoot \,calculation which we have to perform
for the terms that includes the
next powers of $\Fi$. After expanding in derivatives one obtains

\begin{multline}
Z_2[\Fi]= -\int d^{p+1} X \, \lb \tr \d_{k} \Fi_i \d_{l} \Fi_j \,
\, \, \oint \pdt{2} \la x^{k}(t_1) x^{l}(t_2) \ra \la
\st{x}^i(t_1) \st{x} ^j (t_2)\ra + \right .\\
\left . \tr \Fi_i \Fi_j \oint \pdt{2} \la  \st{x}^i(t_1) \st{x} ^j (t_2)\ra  \rb =
  - 2\ap^2 \pi^2   \left (1 - \f 1 \e \right) \int d^{p+1} X \, \tr \d_{k} \Fi_i \d_{k} \Fi_i \,
\end{multline}

Here we see that the mass term also vanishes, and, that
the finite part of the kinetic term coincides with the one
in~(\ref{fip}). Therefore, in this example, the regularization by
analytical continuation is equivalent to dropping the $\f 1 \e$
term (see The Appendix for more details about the regularization).

Then, let us calculate the $\Fi^4$-term by the same expansion near
zero momentum. In this case, the term for constant~$\Fi$-term is equal to
(see The Appendix for more details)
\begin{equation}
Z_4[\Fi]= - \ap^2 \pi^2 \lb 1-\f 1 \e \rb  \int
d^{\PP} X \, \tr [\Fi_i,\Fi_j][\Fi_i,\Fi_j]
\end{equation}
After turning on the field $A$, we shall also get the following expression for the kinetic energy
with derivatives substituted by the covariant ones
\begin{equation}
Z_{2A}[\Fi,A]= 2\ap^2 \pi^2   \left (1 - \f 1 \e \right) \int
d^{\PP} X \, \tr [\d_{k}+A_k,i \Fi_i] [ \d_{k}+A_k,i \Fi_i] \,
\end{equation}
Meanwhile, the kinetic energy of $A$ itself is
\begin{equation}
Z_{A\,{\rm kin}}[A]= - \ap^2 \pi^2   \left (1 - \f 1 \e \right) \int d^{\PP}
X \, \tr F_{kl} F_{kl} \,
\end{equation}
where, as usual $F_{kl} = [\nabla_k, \nabla_l] = [\d_k + A_k, \d_l + A_l]$.

We can combine these terms
\begin{equation}
Z[A,\Fi] = 1 -(\pi\ap)^2\left (1 - \f 1 \e \right)  [\nabla_{\mu},\nabla_{\nu}]^2
\end{equation}
where we denoted
\begin{equation}
\begin{aligned}
 \nabla_{\mu} = \d_{\mu} + A_{\mu}, \qquad \mu = 0\cdots p \\
 \nabla_{\mu} = -i \Fi_{\mu}, \qquad \mu =p+1 \cdots D-1
\end{aligned}
\end{equation}

The divergences of all these terms is the same:
$\sum_{n=1}^{\infty} 1$, and after the regularization (see
The Appendix) by analytical continuation it reduces to
$\sum_{n=1}^{\infty} 1 = -1/2$. The origin of these divergences,
as was stated before, is the tachyon contribution to the exchange diagram. However, after analytical
continuation in momentum it disappears similar to analytically continued
(\ref{prop}) evaluated at $p=0$ with $\ap m^2 =1$.
This statement will be confirmed further by the
calculations in the case of superstring~\cite{Andreev}, where the tachyon
is decoupled, and no linear divergence appears.
We shall encounter the logarithmic divergence
for the first time in $\Fi^6$-term, because
there are non-vanishing vertices
\bfoot
{$-\ap^2 \pi^2\tr[\Fi_i,\Fi_j][\Fi_i,\Fi_j]$}
\efoot
with four legs for constant $\Fi$, and, therefore, there is an infinite at zero momentum diagram
\FiPic, in which the  $\Fi$ field propagates between
these two vertices.

Let us calculate the $\Fi^6$-term. The details of the calculation can
be found in the Appendix at the end of the paper, see also the papers ~\cite{Phi6,Brecher,Dorn}.
The answer includes two structures

\begin{multline}
Z_6[\Fi]= (2\ap)^3 \Big{(} -\f 1 2 A_b + 2 C_b \Big{)} \int d^{p+1} X \, \tr [\Fi_i,\Fi_j][\Fi_j,\Fi_k][\Fi_k,\Fi_i]-\\
- \f 1 4 (2\ap)^3 \Big{(} 2 A_b + 2 B_b \Big{)} \int d^{\PP} X \, \tr
[[\Fi_i,\Fi_j],\Fi_k][[\Fi_i,\Fi_j],\Fi_k]
\end{multline}
Here we denoted
\begin{align}
A_b &\equiv \sum_{n=1}^{\infty} {\pi^2 \f 1 n e^{-3n\e}}\\
B_b &\equiv  \sum_{n,k=1,n\neq k}^{\infty} { {\pi^2 \f 1 n} e^{-(n+2k)\e} }\\
C_b &\equiv  \sum_{n,k=1,n\neq k}^{\infty} { \pi^2 \f n {(k+n)(k-n)}} e^{-(n+2k)\e}
\end{align}
We see that both the linear and logarithmic divergences are present.

The calculation for superstring shows that the linear divergence
$\f 1 \e$ in the term $-(\pi \ap)^2 \lb 1 - \f 1 \e \rb
[\nabla_{\mu},\nabla_{\nu}]$ is cancelled by fermionic
contributions (in the superstring case the tachyon, that causes the linear divergence, is decoupled), and
the final expression is equal exactly to the bosonic one, regularized by analytical
continuation in momentum. But at the $\Fi^6$-term $\ln \f 1 \e$ remains, which is the simple pole $\f
1 {q^2}$ at $q=0$ in terms of the momentum representation. The
expression for $Z_6[\Fi]$ for the superstring is
\begin{multline}
Z_6[\Fi]= (2\ap)^3 \Big{(} -\f 1 2 A_b + 2 C_b  + \f 1 2 A_f - 2 C_f \Big{)} \int d^{p+1} X \, \tr [\Fi_i,\Fi_j][\Fi_j,\Fi_k][\Fi_k,\Fi_i]-\\
- \f 1 4 (2\ap)^3 \Big{(} 2 A_b + 2 B_b - 2 A_f - 2 B_f \Big{)} \int
d^{p+1} X \, \tr [[\Fi_i,\Fi_j],\Fi_k][[\Fi_i,\Fi_j],\Fi_k]
\end{multline}
Here
\begin{align}
A_f &\equiv  \sum_{n=1/2}^{\infty} {\pi^2 \f 1 n e^{-3n\e}}\\
B_f &\equiv \sum_{n,k=1/2,n\neq k}^{\infty} { {\pi^2 \f 1 n} e^{-(n+2k)\e} }\\
C_f &\equiv  \sum_{n,k=1/2,n\neq k}^{\infty} { \pi^2 \f n {(k+n)(k-n)}} e^{-(n+2k)\e}
\end{align}
and
\begin{align}
-\f 1 2 (A_f-A_b) + 2 (C_f-C_b) &\sim - \pi^2 \ln \f 1 \e  \\
2 A_b + 2 B_b - ( 2 A_f + 2 B_f) &\sim -\pi^2 \ln \f 1 \e
\end{align}
Finally
\begin{equation} Z_{6 \rm sstr}= (2\ap)^3 \pi^2 \ln \f 1 \e   \int
d^{\PP} X \, \tr \lb [\Fi_i \Fi_j] [\Fi_j \Fi_k] [\Fi_k \Fi_i] +
\f 1 4 [[\Fi_i \Fi_j],\Fi_k] [[\Fi_i \Fi_j],\Fi_k] \rb \end{equation}

Thus, there is only the logarithmic divergence that has been predicted
before.
It is worth noting that the answer for the partition function at constant $\Phi$ can be expressed in terms of
only commutators of $\Phi^i$. That has a simple interpretation. Indeed, in the abelian case
the $P$-ordering in the partition function $Z[\Phi]$ (\ref{Def_Z_FI}) is not essential, the functional integral can
be easily taken, and the answer is $Z[\Phi] = 1$. Also, in the $T$-dual picture, $\Phi^i$ should be
replaced by the operator $\nabla_i = \d_i + A_i$, however we know that every
symbol $\nabla_i$ should be placed under commutator.
\bfoot
{Otherwise
the overall expression is to be a differential operator on the field functional space, which it should not be.}
\efoot

\section{The effective action for the $\Fi$ fields}
Now we need to get the effective action from the partition function. In the superstring case,
it is
\begin{eqnarray} Z_{int}[\Phi] = -\f 1 4 (2\ap\pi)^2 \tr [\Fi_i,\Fi_j]^2 +\\
+(2\ap)^3 \pi^2 \ln \f 1 \e \tr \lb [\Fi_i \Fi_j] [\Fi_j \Fi_k]
[\Fi_k \Fi_i] + \f 1 4 [[\Fi_i \Fi_j],\Fi_k] [[\Fi_i\Fi_j],\Fi_k]\rb \label {sact}
\end{eqnarray}
We calculated $Z_{int}[\Phi]$ pertubatively in momentum nearby zero, the kinetic term was
\begin{equation}
\f 1 2 \phi K \phi = \f 1 2 (2 \ap \pi)^2 q^2 \tr \Fi(q) \Fi(-q)
\end{equation}
Therefore, we should have $\left . K^{-1} = \infty \right | _{q^2 = 0 }$,
however after regularization this infinity is replaced by $\ln \f 1
\e$. As was stated before,
 from the first quantized field theory we have for the propagator of the massless particle:
\begin{equation}
\int_{0}^{\infty} e^{- \ap q^2 T} \, dT = \f 1 {\ap q^2} \label {ftrepr}
\end{equation}
The regularization by $\e$ in the string theory world
sheet calculations effectively corresponds to restricting the region
of integration in~(\ref{ftrepr}) by $\ln \f 1 \e$
\begin{equation} \left . \f
1 {2 \ap q^2} \right | _{q^2 = 0 } \leftrightarrow \ln \f 1 \e \end{equation}
Thus,
\begin{equation}
\left . K^{-1} \right | _{q^2 = 0}  = \left . \f 1
{(2\ap \pi)^2}  \f 1 {q^2} \right | _{q^2 = 0 }  \leftrightarrow
\f 1 {2\ap \pi^2} \ln \f 1 \e \end{equation}
Now, by using the following
identities
\begin{equation}
\tr \lb [\Fi_i \Fi_j] [\Fi_j \Fi_k] [\Fi_k \Fi_i]
+ \f 1 4 [[\Fi_i \Fi_j],\Fi_k] [[\Fi_i \Fi_j],\Fi_k] \rb = \f 1 2
\tr [[\Fi_i \Fi_j],\Fi_i] [[\Fi_k \Fi_j],\Fi_k]
\end{equation}
and
\begin{equation} \f \de
{\de \Fi_{i}} \tr [\Fi_k,\Fi_j]^2 = 4 [[\Fi_j,\Fi_i],\Fi_j]
\end{equation}
it
can be easily seen that the term
$\f 1 2 \f {\de V} {\de \phi} K^{-1} \f {\de V} {\de \phi}$
in (\ref{zrel}) is exactly equal to
$$(2\ap)^3 \pi^2 \ln \f 1 \e \tr \lb [\Fi_i \Fi_j] [\Fi_j \Fi_k] [\Fi_k \Fi_i] + \f 1 4
[[\Fi_i \Fi_j],\Fi_k] [[\Fi_i \Fi_j],\Fi_k]\rb$$
in (\ref{sact}).

Thus, after renormalization,  the potential in the superstring case, up to the sixth
order in $\Fi$ is equal to
\begin{equation}
S[\Fi] = -\f 1 4 (2\ap\pi)^2 \tr [\Fi_i,\Fi_j]^2 + c \tr [[\Fi_i \Fi_j],\Fi_i] [[\Fi_k \Fi_j],\Fi_k]
\label{fans}
\end{equation}
In this order we can choose $c$ equal to 0. Indeed, making the following field redefinition
\begin{eqnarray}
\Fi_i \to \Fi_i + c_1[[\Fi_k,\Fi_i],\Fi_k] \\
\tr[\Fi_i,\Fi_j]^2 \to \tr ([\Fi_i,\Fi_j]^2 + 4c_1 [[\Fi_i \Fi_j],\Fi_i] [[\Fi_k \Fi_j],\Fi_k]
+ o(\Fi^6)) \label{trans}
\end{eqnarray}
we obtain the same $c$-term structure from the term $\tr[\Fi_i,\Fi_j]^2$, thus $c \to c -(2\ap \pi)^2 c_1$.
This field redefinition also affects the terms containing  higher powers of $\Phi$, but unless
we fix coefficients in the terms of the order more than $\Fi^6$, $c$ can be set to 0.

Also we see in (\ref{fans}), that the structure $\tr[\Fi_i,\Fi_j]^3$ corresponding
in the T-dual picture to the term $\tr F^3$ is absent,
which agrees with the 3-point superstring amplitude for a gauge field
and the papers \cite{tseytlin,Phi6,Brecher,behrndt}
\bigskip

Now let us get the action from the partition function of the bosonic string.
In this case, we have for the partition function the following
expression
\begin{equation}
Z[\Fi] = Z_4[\Fi] + Z_6[\Fi]
\end{equation}
where
\begin{equation}
Z_4[\Fi] = - \ap^2 \pi^2 \lb 1-\f 1 \e \rb  \tr
[\Fi_i,\Fi_j][\Fi_i,\Fi_j]
\end{equation}
and
\begin{multline}
Z_6[\Fi]= (2\ap)^3 \lb \Big{(} -\f 1 2 A_b + 2 C_b \Big{)} \tr
[\Fi_i,\Fi_j][\Fi_j,\Fi_k][\Fi_k,\Fi_i]- \right.\\
- \left. \f 1 4  \Big{(} 2 A_b + 2 B_b
\Big{)}  \tr [[\Fi_i,\Fi_j],\Fi_k][[\Fi_i,\Fi_j],\Fi_k] \rb
\end{multline}

Now, the situation is more subtle than in the superstring
calculations, since the bosonic partition function has not only
logarithmic but also linear divergencies, that should be
regularized by analytical continuation in momentum before
deriving  the action from it. However, we have only the leading
divergent term of the perturbative  expression for $Z_6[\Fi]$ nearby
$q=0$ and naively cannot restore analytically continued expression
from that region of $q$, where it converges. In addition, on-shell, (we
calculated for constant $\Fi$), there is the ${\rm Vol}(SL(2,R))$ (Mobius) infinity.
Fixing $SL(2,R)$ makes finite some terms in $Z_4[\Fi]+Z_6[\Fi]$, but not all of them,
(even in the $\Fi^4$-term, which does not contain contributions  of  the $\Fi$ exchange diagram,
but does contain the tachyon exchange diagram).
However, we can still indirectly restore (regularize) some divergent terms with the $SL(2,R)$ being fixed,
using general additional properties of the potential for constant $\Fi$:
it should not change under the following transformation~\bfoot {It means that it should be constructed only from
commutators of $\Fi^i$.}\efoot:
\begin{equation}
\Fi^i \rightarrow \Fi^i + \Lambda^i, \qquad {\rm if} \qquad [\Lambda^i,\Lambda^j]=0,\qquad [\Lambda^i,\Fi^j]=0
\end{equation}
and should be the trace of polynomial of $\Fi^i$.

Thus, there are only one structure for $\Fi^4$-term
\begin{equation}
\tr [\Fi_i,\Fi_j]^2
\end{equation}
and only two linear independent structures for
$\Fi^6$-term:
\begin{align}
Z_{6_1}=\tr [\Fi_i \Fi_j] [\Fi_j \Fi_k] [\Fi_k \Fi_i]\\
Z_{6_2}=\tr [[\Fi_i \Fi_j],\Fi_i] [[\Fi_k \Fi_j],\Fi_k]
\end{align}

As usual, we can omit integration over $SL(2,R)$ by arbitrarily fixing three points,
multiplying the integrand by the Faddev-Popov determinant for $SL(2,R)$ on the  unit circle
\bfoot
{The norm has been chosen in such a way that the $SL(2,R)$ fixing for $\Fi^4$-term corresponds
to the previous regularization by $\zeta$-function.}
\efoot
\begin{equation}
\Delta_{FP} = (4\pi)^2 \sin \lb \f {\phi_2-\phi_1} 2 \rb \sin \lb
\f {\phi_3-\phi_2} 2\rb \sin \lb \f {\phi_1-\phi_3} 2\rb,
\end{equation}
integrating over the positions of remaining points, and dividing the
result by the number of external legs
(since in the initial region of integration the first point is always
the nearest to zero, but the $SL(2,R)$ transformation can cyclically permute points).
In such a way we, one  gets the finite expression for this contribution in the order $\Fi^4$
\begin{equation}
Z_4[\Fi] \supset -2(\pi \ap)^2 \tr \Fi_i \Fi_j \Fi_i \Fi_j
\end{equation}
and we can uniquely restore the full expression for
$Z_4[\Fi]=-\f 1 4 (2 \pi \ap)^2 \tr [\Fi_i,\Fi_j][\Fi_i,\Fi_j]$.
Similarly, there is the finite answer for the coefficient of
the $\tr \Fi_i \Fi_j \Fi_k \Fi_i \Fi_j \Fi_k$ term in order $\Fi^6$
\begin{equation}
Z_6[\Fi] \supset  \f {(2\ap)^3 \pi^2} 3 \tr \Fi_i \Fi_j \Fi_k \Fi_i \Fi_j \Fi_k
\end{equation}
However, we know that the structure $Z_{6_2}$
\begin{equation}
\tr [[\Fi_i \Fi_j],\Fi_i] [[\Fi_k \Fi_j],\Fi_k] \label{z62}
\end{equation}
does not contain
\begin{equation}
\tr \Fi_i \Fi_j \Fi_k \Fi_i \Fi_j \Fi_k,
\end{equation}
i.e. $\tr \Fi_i \Fi_j \Fi_k \Fi_i \Fi_j \Fi_k$
can only emerge from $Z_{6_1}$
\begin{equation}
\tr [\Fi_i \Fi_j] [\Fi_j \Fi_k] [\Fi_k \Fi_i] \supset - \tr \Fi_i \Fi_j \Fi_k \Fi_i \Fi_j \Fi_k
\end{equation}

Now we almost know the answer for
the full $Z_6[\Fi]$:
\begin{equation}
Z_6[\Fi] = - \f {(2\ap)^3 \pi^2} 3 \tr [\Fi_i \Fi_j] [\Fi_j \Fi_k] [\Fi_k \Fi_i] + \lambda Z_{6_2},
\end{equation}
where $\lambda$ is yet arbitrary. However, the structure $Z_{6_2}$
appears in the partition function from the $\Fi$ exchange diagram
(i.e. from the term $\f 1 2 \f {\de V} {\de \phi} K^{-1} \f {\de
V} {\de \phi}$ with $V \sim [\Fi_i,\Fi_j]^2$ and should be subtracted to get the action.

Thus, the potential in the bosonic string case,
up to the sixth order in $\Fi$ is equal to
\begin{equation}
S[\Fi]= -\f 1 4 (2\ap\pi)^2 \tr [\Fi_i,\Fi_j]^2 -
\f {(2\ap)^3 \pi^2} 3 \tr [\Fi_i \Fi_j] [\Fi_j \Fi_k] [\Fi_k \Fi_i] +
c [[\Fi_i \Fi_j],\Fi_i] [[\Fi_k \Fi_j],\Fi_k], \label{bans}
\end{equation}
where $c$ can be set to $0$ unless we fix coefficients of the terms of higher powers in $\Fi$.
(The reason is the same as in the superstring case, see above.)

The structure $\tr[\Fi_i,\Fi_j]^3$ in (\ref{bans}), corresponds to
the $\tr F^3$-term in the T-dual picture, see the papers \cite{tseytlin,scherk,Brecher}.

\section{Conclusion}
We discussed the connection between the partition function for the $2D$ sigma model with boundary
pertubations and the tree level effective action for massless fields in the open string theory.
By direct calculation we obtained the potential for the non-abelian massless scalars
on a stack of $N$ D$p$-brane up to the order $\Fi^6$.
In the case of bosonic the string, the answer is
\begin{equation}
V[\Fi]= -\f 1 4 (2\ap\pi)^2 \tr [\Fi_i,\Fi_j]^2 - \f {(2\ap)^3 \pi^2} 3 \tr [\Fi_i \Fi_j]
[\Fi_j \Fi_k] [\Fi_k \Fi_i] + O(\Fi^8)
\end{equation}
In the case of the superstring, the answer is
\begin{equation}
V[\Fi] = -\f 1 4 (2\ap\pi)^2 \tr [\Fi_i,\Fi_j]^2 + O(\Fi^8)
\end{equation}
There is another structure (\ref{z62}) that can arise in the order $\Fi^6$ (\ref{fans},\ref{bans}), but its coefficient
can be made arbitrary by field redefinition, see (\ref{trans}).

\section{Acknowledgments}
Author is grateful to A.Mironov for careful reading the manuscript,
to A.Dymarsky, A.Alexandrov, V.Dolgushev and B.Kachura for useful discussions and especially to
E.Akhmedov and A.Morozov for initiating this work and support.
This work was partly supported by the Russian President's grant 00-15-99296,  RFBR grant 01-02-17488 and
INTAS grant 00-00561.

\appendix

\section{Appendix}
\bigskip

For evaluation of the partition function~(\ref{pf}) we need
the 2-dimensional propagators for $X^{\mu}$ with the Dirichlet and Neumann boundary
conditions. They are equal to the Green functions of the Laplace operator
on the disk with appropriate boundary conditions, since
\begin{equation}
-\f 1 {4\pi\ap} \int {\d_a x_{\mu} \d^a x^{\mu} } \, d^2\sigma
= \f 1 {4\pi\ap} \int {x_{\mu} \Delta x^{\mu} } \, d^2\sigma  =
-\f 1 2 \int {x^{\mu} G^{-1}_{\mu\nu} x^{\nu} } \, d^2\sigma
\end{equation}
Here $x(\s)$ does not contain the zero mode
$$ \int x(\s) \, d^2\sigma = 0$$

On the complex plane $z = \s_1 + i\s_2$ with unspecified boundary conditions the propagator is
\begin{align}
 &G(z_1,z_2)_{\mu\nu} = g_{\mu\nu}G(z_1,z_2)\\
 &G(z_1,z_2) = -\ap \ln |z_1-z_2|
\end{align}
The propagators on the flat unit disk with the Dirichlet or Neumann boundary conditions could be found by the method
of images,
\begin{align}
G_d(z_1,z_2) &= -\ap \ln \left | \f {z_1-z_2} {1-z_1\bar{z_2}}\right | \label{gd} \\
G_n(z_1,z_2) &= -\ap \ln \left | 4 (z_1-z_2)
(1-z_1\bar{z_2})\right |   \label{gn}
\end{align}
We will also need the propagator derivatives on the boundary.
Here $z_1=r_1 e^{it_1}, \enskip z_2=r_2 e^{it_2}$. On the disk
boundary ($r_1=r_2=1$) we have for the Dirichlet boundary conditions ($\dt x^i = 0$)
\begin{align}
&\la x^i(z_1) x^j(z_2) \ra = g^{ij}G_d(z_1,z_2) = 0 \\
&\la \st{x}^i(z_1) x^j(z_2) \ra =g^{ij} \d_{r_1} G_d(z_1,z_2) = 0 \\
&\la \st{x}^i(z_1) \st{x}^j(z_2) \ra =g^{ij} \d_{r_1} \d_{r_2}
G_d(z_1,z_2) = g^{ij} \ap \f  1 {2 \sin^2 \left ( \f {t_1 - t_2} 2
\right )} &=
- 2g^{ij} \, \ap \sum_{n=1}^{\infty} { e^{-n \e} n  \cos {n(t_1 - t_2)}}
\end{align}
and for the Neumann boundary conditions  ($\dn x^k = 0$)
\begin{equation}
\la x^k(z_1) x^l(z_2) \ra =g^{kl}\,G_n(z_1,z_2) = - g^{kl}\, \ap \ln \left | 4 \sin^2
\left ( \f {t_1 - t_2} 2  \right ) \right | = 2 g^{kl}\, \ap \sum_{n=1}^{\infty}
{e^{-n \e} \f 1 n \cos {n(t_1 - t_2)}}
\end{equation}
\begin{equation}
\la \t{x}^k(z_1) x^l(z_2) \ra =g^{kl}\,\d_{t_1} G_n(z_1,z_2) = - g^{kl}\, \ap \ctg
\left ( \f {t_1 - t_2} 2  \right ) = -2 g^{kl}\, \ap \sum_{n=1}^{\infty} { e^{-n \e} \sin {n(t_1 - t_2)}}
\end{equation}
\begin{equation}
\la \t{x}^k(z_1) \t{x}^l(z_2) \ra =g^{kl}\,\d_{t_1}
\d_{t_2}G_n(z_1,z_2) = - g^{kl}\, \ap \f  1 {2\sin^2 \left ( \f
{t_1 - t_2} 2  \right )} =\phantom{-\,\,}2 g^{kl}\, \ap
\sum_{n=1}^{\infty} { e^{-n \e} n \cos {n(t_1 - t_2)}}
\end{equation}

Regularization by multiplying terms in the sums by $e^{-n \e}$ is
equivalent to dropping out high modes of the restricted to the
world-sheet boundary Laplace operator; its a cutoff at small
distance $\e$ on the world sheet.

\subsection{Calculations of $\Fi^2$-term}

\begin{multline}
Z_2[\Fi]=-\int d^{p+1} X\la \oint \pdt{2} \tr \Fi_i(X+ x(t_1)) \Fi_j(X+ x(t_2)) \st{x}^i(t_1) \st{x}^j(t_2)  \ra = \\
= -\sum_{q_1,q_2} (2\pi)^{p+1} \delta (q_1 + q_2) \tr \Fi_i(q_1)
\Fi_j(q_2) \la \oint \pdt{2} e^{iq_1 x (t_1)} e^{iq_2 x(t_2)}
\st{x}^i(t_1) \st{x}^j(t_2) \ra
\end{multline}
For the Gaussian integrals we have \bfoot{ The measure $\D x$ is taken so, that $\int \D x \,
e^{-\f 1 2 x G^{-1} x} = 1$} \efoot
$$ \int \D x \,e^{-\f 1 2 x G^{-1} x + i J x} = e^{ - \f 1 2 J G J}$$
Correlators are
\begin{align}
&\la \st{x}^i(t_1) \st{x}^j(t_2) \ra = g^{ij} \d_{r_1} \d_{r_2}
G_d(t,t') = g^{ij} \ap \f  1 {2 \sin^2 \left ( \f {t_1 - t_2} 2
\right )}  =
g^{ij} \, (-2 \ap)  \sum_{n=1}^{\infty} e^{-n\e} n \cos n(t_1 - t_2 ) \\
&\la x^k(t_1) x^l(t_2) \ra = -g^{kl} \, \ap \ln \left | 4 \sin^2
\left ( \f {t_1 - t_2} 2  \right ) \right | = g^{kl} \, (2 \ap)
\sum_{n=1}^{\infty} e^{-n\e} \f 1 n  \cos n(t_1 - t_2 )
\end{align}
Thus,
\begin{multline}
\la \oint \pdt{2} e^{ip_1 x (t_1)} e^{ip_2 x(t_2)} \st{x}^i(t_1)
\st{x}^j(t_2) \ra  =
 g^{ij} \oint  \pdt{2}  \lb 4 \sin^2 \lb \f {t_1-t_2} 2 \rb \rb ^ {\ap q_1 q_2} \f \ap {2 \sin^2 \lb \f {t_1-t_2} 2 \rb  } =\\
 = g^{ij} \, 2 \ap \, \pi \int_{0}^{2\pi}    \lb 4 \sin^2 \lb \f t 2 \rb \rb ^ {\ap q_1 q_2 - 1 }  \,dt=
 g^{ij} \,\,2 \ap \,\, 4^{\ap q_1 q_2 -1}  \f {\G(\ap q_1 q_2 - \f 1 2)} { \G(\ap q_1 q_2 )} \,\,\f {2\pi^2 } { \sqrt{\pi}}
\end{multline}
Here we used the regularization by analytical continuation in
momentum.

The expansion in derivatives is
\begin{multline}
Z_2[\Fi]= -\int d^{p+1} X \, \lb \tr \d_{k} \Fi_i \d_{l} \Fi_j \,
\, \, \oint \pdt{2} \la x^{k}(t_1) x^{l}(t_2) \ra \la
\st{x}^i(t_1) \st{x} ^j (t_2)\ra \right.+ \\ \left.
+\tr \Fi_i \Fi_j \oint \pdt{2} \la  \st{x}^i(t_1) \st{x} ^j (t_2)\ra  \rb
= -A_1 \int  \tr \Fi_i \Fi_i \, d^{p+1} X  -A_2 \int  \tr \d_k
\Fi_i \d_k \Fi_i \, d^{p+1} X
\end{multline}
where
\begin{equation}
 A_1 = \oint \pdt{2} (-2 \ap) \lb \sum_{n=1}^{\infty} e^{-n\e} n \cos n(t_1 - t_2 ) \rb =0
\end{equation}
\begin{multline}
 A_2 = \oint \pdt{2} (-2 \ap) \lb \sum_{n=1}^{\infty} e^{-n\e} n \cos n(t_1 - t_2 ) \rb  (2 \ap) \lb \sum_{n=1}^{\infty} e^{-n\e} \f 1 n \cos n(t_1 - t_2 ) \rb =\\
 =-4\pi^2 \ap^2 \sum_{n=1}^{\infty} e^{-2 n\e }=2\pi^2\ap^2 \lb 1 - \f 1 \e \rb
\end{multline}
Then,
\begin{equation}
Z_2[\Fi] =  - 2 \ap^2 \pi ^2 \lb 1 - \f 1 {\e} \rb
\int d^{\PP} X \, \tr \d_k \Fi_i \d_k \Fi_i \label{eps}
\end{equation}

We see, that $\f 1 \e$ term should be dropped to make the both
regularizations equivalent (i.e.~(\ref{eps}) coinciding
with~(\ref{fip})). It is also equivalent to the regularization by
the $\zeta$-function, since $\sum_{n=1}^{\infty} 1 \req \zeta(0) = -\f
1 2 $.

In addition, one can regularize divergent integrals $A_1$, $A_2$
by analytical continuation around the poles  at
$t = 2\pi n, \enskip n \in \mathbb{Z}$

\begin{multline}
A_1 = \oint \pdt{2}  \f \ap {2 \sin^2 \lb \f {t_1 - t_2} 2 \rb } =
\ap \pi \int_{0}^{2\pi} \f 1 {2 \sin^2 \lb \f t 2 \rb } \, dt \req
\ap \pi \int_{0+2i\xi}^{2\pi+2i\xi} \f 1 {2 \sin^2 \lb \f t 2 \rb
} \, dt \req \\ \req  \ap \pi (\ctg (\xi) - \ctg(\xi + \pi)) = 0
\end{multline}
Thus, $A_1$ vanishes after analytical continuation due to
$\pi$-periodicity of $\ctg(t)$.

Let us see, what happens with $A_2$
\begin{multline}
A_2 \req \oint \pdt{2}  \f \ap {2 \sin^2 \lb \f {t_1 - t_2} 2 \rb
} (-\ap) \ln \left | 4 \sin^2 \left ( \f {t_1 - t_2} 2  \right )
\right |
\req -\pi \ap^2 \int_{2\xi}^{2\pi + 2\xi} \f 1 {2 \sin^2 \f t 2 }\, \ln \sin^2 \f t 2 \,\, dt \req \\
\req -\pi \ap^2 \int_{2\xi}^{2\pi + 2\xi} \ctg^2 {\f t 2} \, \, dt
\req -\pi \ap^2 \int_{2\xi}^{2\pi + 2\xi} \lb \f 1 {\sin^2 \f t 2
} - 1  \rb\,dt \req 2\pi^2\ap^2
\end{multline}

We see that this regularization coincides with the two previous ones,
the analytical continuation in momentum and the regularization of
$\sum_{n=0}^{\infty} 1 $ by $\zeta(0)$ (or dropping $\f 1 \e$,
after changing $\sum_{n=0}^{\infty} 1$ to $\sum_{n=0}^{\infty}
e^{-n \e}$).

\subsection {Calculations of $\Fi^4$-term}

By $\zz(\Fi)$ we denote the local density of the partition
function $Z[\Fi]$,
\begin{equation}
Z[\Fi] = \int d^{\PP} X \, \zz(\Fi)
\end{equation}

\begin{multline}
\zz_{4}(\Fi) =\tr \int \limits_{0}^{2\pi} dt_4  \int
\limits_{0}^{t_4} dt_3  \int \limits_{0}^{t_3} dt_2  \int
\limits_{0}^{t_2} dt_1
\Fi_i \Fi_j \Fi_k \Fi_l \la \d_n X^{i}(t_1) \d_n X^{j} (t_2) \d_n X^{k} (t_3)\d_n X^{l} (t_4)\ra = \\
=\tr \Fi_i \Fi_j \Fi_k \Fi_l \int \limits_{0}^{2\pi} dt_4  \int
\limits_{0}^{t_4} dt_3  \int \limits_{0}^{t_3} dt_2  \int
\limits_{0}^{t_2} dt_1
 ( \de^{ij} \de^{kl} G''(t_1,t_2)G''(t_3,t_4)  + \\ + \de^{ik} \de^{jl} G''(t_1,t_3)G''(t_2,t_4) + \de^{il} \de^{jk} G''(t_1,t_4)G''(t_2,t_3) )= \\
 = (2\ap)^2 \tr \sum_{n,m=1}^{\infty} \left(\Fi_i \Fi_i \Fi_j \Fi_j \cdot 0  + (\Fi_i \Fi_j \Fi_i \Fi_j - \Fi_i \Fi_j \Fi_j \Fi_i)
e^{-n\e-m\e}\de_{nm} \cdot n \, m \cdot \f {\pi^2} {n^2}\right) =\\
= (2 \pi \ap ) ^2 \tr \f 1 2 [\Fi_i, \Fi_j]^2 \sum_{n=1}^{\infty} e^{-2n\e} = \\
= (2 \pi \ap )^2 \tr \f 1 2 [\Fi_i, \Fi_j]^2 \left (\f 1 {2\e} -
\f 1 2 + O(\e) \right) = -(\pi \ap)^2 \tr [\Fi_i, \Fi_j]^2 \lb 1 -
\f 1 \e \rb
 \end{multline}
\bigskip

Now let us calculate  $Z_4[\Fi]$ for the superstring case.
The partition function for the superstring is
\begin{equation}
\zz(\Fi) = \la \tr \pexp \left\{\oint \left(i \Fi_i \dt X^i + \f 1 2
[\Fi_i,\Fi_j]\psi_i\psi_j \,\right)d\tau    \right\} \ra
\end{equation}
Averaging is done with
\begin{equation}
\la 1 \ra =\int \D X \ \D \psi \,\exp{\left\{  - \f 1 2 \int { (XG_b^{-1}X}+\psi G_f^{-1} \psi ) \,
d^2\sigma \right\}}
\end{equation}
The restriction of the fermionic propagator to the boundary is:
\begin{equation}
G_f(\phi,\phi') = 2\ap \sum_{n=0}^{\infty} {e^{-(n + \f 1 2)\e} \sin (n+1/2)(\phi-\phi') }
\end{equation}

Firstly, let us correct our bosonic calculation.
One should add the following expression due to
the fermionic contribution
\begin{multline}
\zz_{4 ferm}(\Fi) =\lb- \f 1 2 \rb^2 \tr \int \limits_{0}^{2\pi}
dt_2  \int \limits_{0}^{t_2} dt_1
[\Fi_i \Fi_j][ \Fi_k \Fi_l] \la \psi^i(t_1) \psi^j(t_1) \psi^k(t_2) \psi^l(t_2) \ra=\\
=- \lb - \f 1 2  \rb ^2 \lb (2\ap)^2 \pi^2 \tr \lb [\Fi_i \Fi_j][
\Fi_i \Fi_j]-[\Fi_i \Fi_j][ \Fi_j \Fi_i] \rb \rb
\lb \sum_{n=0}^{\infty}{e^{-(n + \f 1 2)\e}}\rb \\
= - (\ap \pi)^2 2 \tr [\Fi_i \Fi_j]^2 \lb \f 1 {2\e}+O(\e) \rb
\end{multline}

Now the fermionic term cancels the divergence in the bosonic term.
\begin{equation}
\zz_{4} = - (\ap \pi)^2 \tr [\Fi_i \Fi_j]^2 \lb 1 - \f 1 {\e}+ \f
1 {\e} + O(\e) \rb = - (\ap \pi)^2 \tr [\Fi_i \Fi_j]^2
\end{equation}

\subsection {Calculations of $\Phi^6$-term}

We introduce the notations:
\begin{align}
A_b &\equiv \sum_{n=1}^{\infty} {\pi^2 \f 1 n e^{-3n\e}}\\
B_b &\equiv \sum_{n,k=1,n\neq k}^{\infty} { {\pi^2 \f 1 n} e^{-(n+2k)\e} }\\
C_b &\equiv \sum_{n,k=1,n\neq k}^{\infty} { \pi^2 \f n {(k+n)(k-n)}} e^{-(n+2k)\e} \\
D_b &\equiv \sum_{n,k=1,n\neq k}^{\infty} { \pi^2 \f {k^2} {n(k+n)(k-n)}} e^{-(n+2k)\e} \\
A_f &\equiv \sum_{n=1/2}^{\infty} {\pi^2 \f 1 n e^{-3n\e}}\\
B_f &\equiv \sum_{n,k=1/2,n\neq k}^{\infty} { {\pi^2 \f 1 n} e^{-(n+2k)\e} }\\
C_f &\equiv \sum_{n,k=1/2,n\neq k}^{\infty} { \pi^2 \f n {(k+n)(k-n)}} e^{-(n+2k)\e} \\
D_f &\equiv \sum_{n,k=1/2,n\neq k}^{\infty} { \pi^2 \f {k^2}  {n(k+n)(k-n)}} e^{-(n+2k)\e}
\end{align}

Some of these sums can be expressed through the others,
\begin{align}
D_b - B_b &= C_b \\
D_f - B_f &= C_f
\end{align}

The bosonic part contains 15 terms, corresponding to different pairing
of X.

\begin{multline}
\zz_{6XXXXXX}(\Fi) =\tr    \int \limits_{0}^{2\pi} dt_6 \int
\limits_{0}^{t_6} dt_5 \int \limits_{0}^{t_5} dt_4  \int
\limits_{0}^{t_4} dt_3  \int \limits_{0}^{t_3} dt_2  \int
\limits_{0}^{t_2} dt_1
\Fi_i \Fi_j \Fi_k \Fi_l \Fi_m \Fi_n \\
\la \d_n X^{i}(t_1) \d_n X^{j} (t_2) \d_n X^{k} (t_3)\d_n
X^{l}(t_4) \d_n X^{m}(t_5) \d_n X^{n}(t_6)\ra
\end{multline}

However, due to the cyclic symmetry of the $\Fi$-product under the trace,
only the 5 different structures remain

\begin{equation}
\begin{aligned}
\zz_{XXXXXX}(\Fi) = (2\ap)^3 (
  &\tr \Fi_i \Fi_i \Fi_j \Fi_j \Fi_k \Fi_k  \lb -\f 1 2  A_b + 2 C_b \rb + \\
+ &\tr \Fi_i \Fi_i \Fi_j \Fi_k \Fi_j \Fi_k  \lb \f 7 2 A_b + 2 B_b -6 C_b \rb + \\
+ &\tr \Fi_i \Fi_i \Fi_j \Fi_k \Fi_k \Fi_j  \lb -2 A_b - 2 B_b \rb + \\
+ &\tr \Fi_i \Fi_j \Fi_k \Fi_i \Fi_k \Fi_j  \lb \f 1 2 A_b + 4 C_b +2 D_b \rb +\\
+ &\tr \Fi_i \Fi_j \Fi_k \Fi_i \Fi_j \Fi_k  \lb -\f 3 2 A_b -2 D_b \rb ) = \\
= (2\ap)^3 (&\tr [\Fi_i,\Fi_j][\Fi_j,\Fi_k][\Fi_k,\Fi_i]\lb -\f 1 2 A_b + 2 C_b \rb + \\
+ &\tr [[\Fi_i,\Fi_j],\Fi_k][[\Fi_i,\Fi_j],\Fi_k] \f 1 4 \lb -2
A_b -2 B_b \rb)
\end{aligned}
\end{equation}

In the superstring case also the following terms should be added

\begin{equation}
\begin{aligned}
\zz_{XX\psi\psi\psi\psi}(\Fi) = (2\ap)^3 (
  &\tr \Fi_i \Fi_i \Fi_j \Fi_j \Fi_k \Fi_k  \lb 0 \rb + \\
+ &\tr \Fi_i \Fi_i \Fi_j \Fi_k \Fi_j \Fi_k  \lb -2 A_f -2 B_f \rb + \\
+ &\tr \Fi_i \Fi_i \Fi_j \Fi_k \Fi_k \Fi_j  \lb 2 A_f + 2 B_f \rb + \\
+ &\tr \Fi_i \Fi_j \Fi_k \Fi_i \Fi_k \Fi_j  \lb -2 A_f - 2 B_f \rb +\\
+ &\tr \Fi_i \Fi_j \Fi_k \Fi_i \Fi_j \Fi_k  \lb 2 A_f + 2 B_f \rb ) = \\
=  (2\ap)^3 &\tr [[\Fi_i,\Fi_j],\Fi_k][[\Fi_i,\Fi_j],\Fi_k] \f 1 4
\lb 2 A_f + 2 B_f \rb
\end{aligned}
\end{equation}

and

\begin{equation}
\begin{aligned}
\zz_{\psi\psi\psi\psi\psi\psi}(\Fi) = (2\ap)^3 (
  &\tr \Fi_i \Fi_i \Fi_j \Fi_j \Fi_k \Fi_k  \lb -1(-\f 1 2 A_f + 2 C_f) \rb + \\
+ &\tr \Fi_i \Fi_i \Fi_j \Fi_k \Fi_j \Fi_k  \lb 3 (-\f 1 2 A_f + 2 C_f)\rb + \\
+ &\tr \Fi_i \Fi_i \Fi_j \Fi_k \Fi_k \Fi_j  \lb 0 \rb + \\
+ &\tr \Fi_i \Fi_j \Fi_k \Fi_i \Fi_k \Fi_j  \lb -3 (-\f 1 2 A_f + 2 C_f) \rb +\\
+ &\tr \Fi_i \Fi_j \Fi_k \Fi_i \Fi_j \Fi_k  \lb 1  (-\f 1 2 A_f + 2 C_f)\rb )=\\
= (2\ap)^3 &\tr [\Fi_i,\Fi_j][\Fi_j,\Fi_k][\Fi_k,\Fi_i]\lb \f 1 2 A_f - 2 C_f \rb
\end{aligned}
\end{equation}

Thus, the full expression for the $\Fi^6$-term in the superstring case is

\begin{equation}
\begin{aligned}
\zz_{6 \rm sstr}(\Fi) =
(-2\ap)^3 \left( \right.&\tr [\Fi_i,\Fi_j][\Fi_j,\Fi_k][\Fi_k,\Fi_i]\lb -\f 1 2 A_b + 2 C_b \rb + \\
+ &\tr [[\Fi_i,\Fi_j],\Fi_k][[\Fi_i,\Fi_j],\Fi_k] \f 1 4 \lb -2 A_b -2 B_b  \rb\\
+ &\tr [[\Fi_i,\Fi_j],\Fi_k][[\Fi_i,\Fi_j],\Fi_k] \f 1 4 \lb 2 A_f + 2 B_f \rb\\
+ &\tr [\Fi_i,\Fi_j][\Fi_j,\Fi_k][\Fi_k,\Fi_i]\lb \f 1 2 A_f - 2 C_f \rb )
\end{aligned}
\end{equation}
which reduces to
\begin{equation}
\begin{aligned}
\zz_{6 \rm sstr}=  (2\ap)^3 \pi^2 \ln \f
1 \e  \tr \lb [\Fi_i \Fi_j] [\Fi_j \Fi_k] [\Fi_k \Fi_i] + \f 1 4
[[\Fi_i \Fi_j],\Fi_k] [[\Fi_i \Fi_j],\Fi_k] \rb = \\
=(2\ap)^3 \pi^2 \ln \f
1 \e  \tr \lb  \f 1 2 [[\Fi_i \Fi_j],\Fi_i] [[\Fi_k \Fi_j],\Fi_k] \rb
\end{aligned}
\end{equation}

\end{document}